\documentclass[final,authoryear,5p,times,twocolumn]{elsarticle}
\usepackage{graphicx}

\usepackage{amssymb}

\journal{Radiation Measurements}

\begin{document}

\begin{frontmatter}



\title{Anomalous europium luminescence in LaF$_3$}


\author{E A Radzhabov$^{1,2}$}
\ead{eradzh@igc.irk.ru}

\author{R Yu Shendrik$^{1,2}$}

\address{$^1$Vinogradov Institute of Geochemistry, Russian Academy of Sciences, Favorskii street 1a, P.O.Box 4019, 664033 Irkutsk, Russia}

\address{$^2$Irkutsk State University, Physics department, Gagarin boulevard 20, 664003 Irkutsk, Russia}

\begin{abstract}
Optical spectra (absorption, emission, excitation, decay) and dielectric relaxation were measured for divalent europium (and partially for ytterbium) in lanthanum fluoride crystals. Absorption of Eu$^{2+}$ contains not only asymmetric weakly structured band at 245 nm but also long-wavelength bands at 330, 380 nm. Broadband Eu$^{2+}$ emission at 600 nm appeared below 80 K, having decay time 2.2 $\mu$s at 7.5 K. Emission at 600 nm is attributed to so-called anomalous luminescence. 

Bulk conductivity is directly proportional to absorption coefficient of Eu$^{2+}$ bands. Dielectric relaxation peak of LaF$_3$-EuF$_3$ is attributed to rotation of dipoles Eu$^{2+}$-anion vacancy. The long-wavelength absorption bands at 330, 380 nm are assigned to transitions from 4f$^7$ Eu$^{2+}$ ground state to states of neighbouring fluorine vacancy.

\end{abstract}


\end{frontmatter}

\section{Introduction}
Europium Eu$^{2+}$ ions are known as very efficient luminescence impurity in dense scintillating hosts. Europium introduced into halide crystal in divalent or trivalent states.  A number of investigations are devoted to trivalent lanthanides in LaF$_3$ crystals \citep{carnall1989systematic, heaps1976vacuum}. At the same time authors noted the tendency of  EuF$_{3}$, to reduce to EuF$_{2}$ at the high temperatures required for LaF$_{3}$ crystal growth, and the very strong broad band structure associated with Eu$^{2+}$ in the visible and ultraviolet range due to 4f - 4f5d transitions \citep{carnall1989systematic}. 

Absorption bands of Eu$^{2+}$ were observed at 280 nm in LaCl$_{3}$ \citep{gruen1956color} and at 245 nm in LaF$_3$ \citep{heaps1976vacuum}. Eu$^{2+}$ luminescence was found in LaCl$_3$ crystal at 420 nm \citep{kim1967electron, lehmann1975heterogeneous}. Both absorption and emission are obviously due to transitions between ground 4f$^7$ and excited 4f$^6$5d$^1$ Eu$^{2+}$ states. No data on luminescence of Eu$^{2+}$ in LaF$_{3}$ were found in literature. 

Besides the normal 5d-4f luminescence in most materials, the Eu$^{2+}$ (and also Yb$^{2+}$ and others) show so called "anomalous" broadband luminescence with large Stokes shift in certain crystals (see review \citep{dorenbos2003anomalous}, \citep{grinberg2008impurity}). For such crystals the excited 5d level falls into conduction band. Luminescence observed after transitions from conduction band states, which have less energy than the 5d level, to 4f level of lanthanide impurity ion \citep{moine1989luminescence, dorenbos2003anomalous}. 

Divalent impurity ion has charge less than the charge of lanthanum, therefore for the electrical neutrality of the LaF$_3$ crystal the additional positive charge is needed for each divalent ion. In the absence of oxygen the charge compensation of divalent ion Ca$^{2+}$, Sr$^{2+}$ or Ba$^{2+}$ in LaF$_3$ accomplished by fluorine vacancy \citep{roos1985dielectric, igel1982electrical}. Parallel growth of ac conductivity and absorption  in the visible region were observed in LaF$_3$-Sm$^{2+}$. Conductivity was attributed to Sm$^{2+}$-fluorine vacancy reorientation \citep{radzhabov2015spectra}. The dipoles in solids were thoroughly investigated by dielectric relaxation \citep{jonscher1999dielectric, schonhals2003analysis}. 

The main topic of the present paper is to study the optical and dielectric properties of divalent Eu and Yb in LaF$_3$.

\section{Experimental}

Crystals were grown in vacuum in a graphite crucible by the Stockbarger method \citep{Radzhabov2012}. The graphite crucible contained three cylindrical cavities 10 mm in diameter and 80 mm long, which allowed growing three crystals of {\O}10x50 mm dimensions with different impurity concentrations at the same time. A few percent of CdF$_2$ was added into raw materials for purification from oxygen impurity during growth. Impurity LnF$_3$ (Ln – lanthanide) was added into LaF$_3$ powder in concentration of 0.01, 0.1 and 0.3 mol.\%. In LaF$_3$-YbF$_3$ crystals the Ce$^{3+}$ absorption at 245 nm and less wavelengths was found, which not influenced on Yb$^{2+}$ bands identification. The samples {\O}10 mm x 2mm sawed from the grown rods and polished were typically used for measurements. 

Absorption spectra in the range 190-3000 nm were taken with spectrophotometer Perkin-Elmer Lambda-950, emission spectra were measured using grating monochromator MDR2 (LOMO). Emission, excitation spectra were measured with photomodule Hamamatsu H6780-04 (185-850nm). No emission spectrum correction needs to be performed as the sensitivity only weakly changed in the region of Eu emission (400-700 nm). X-irradiation was performed using Pd-tube with 40 kV 20 mA.

As electrode contact material a silver paint (kontaktol "Kettler") was employed. Diameter of paint electrodes was around 5 mm and crystals thickness was around 2 mm. Conductivity measurements were done using immitance (RLC) meter E7-20 (“MNIPI”) in frequency range 25 Hz- 1MHz. Comparative LaF$_3$ and LaF$_3$-Eu dielectric measurements were done at room temperature. Conductivity of LaF$_3$ could be measured at mono-frequency. However, the bulk conductivity determined in this way appears to be mostly too small \citep{roos1985bulk}. We investigate the conductivity of LaF$_3$-Sm$^{2+}$ in previous paper at frequency 1 kHz \citep{radzhabov2015spectra}. While the  relation between conductivity of samples with different Sm$^{2+}$ concentrations remains the same, the values of measured conductivity were several time smaller. Therefore in this paper we measure true bulk LaF$_3$ conductivity from frequency dispersion \citep{schoonman1980solid}.

\section{Results}

\subsection{Optical spectra}

Eu$^{3+}$ ions easily recognized in LaF$_3$ \citep{carnall1989systematic} and in many other materials by sharp red emission lines due to f-f transitions. No red luminescence due to Eu$^{3+}$ was found in all our LaF$_3$-Eu crystals at 7.5-300 K while intensive ultraviolet absorption appeared. Therefore europium impurity introduced in divalent form in our LaF$_3$ crystals.

\begin{figure} [h]
\includegraphics[width=20pc]{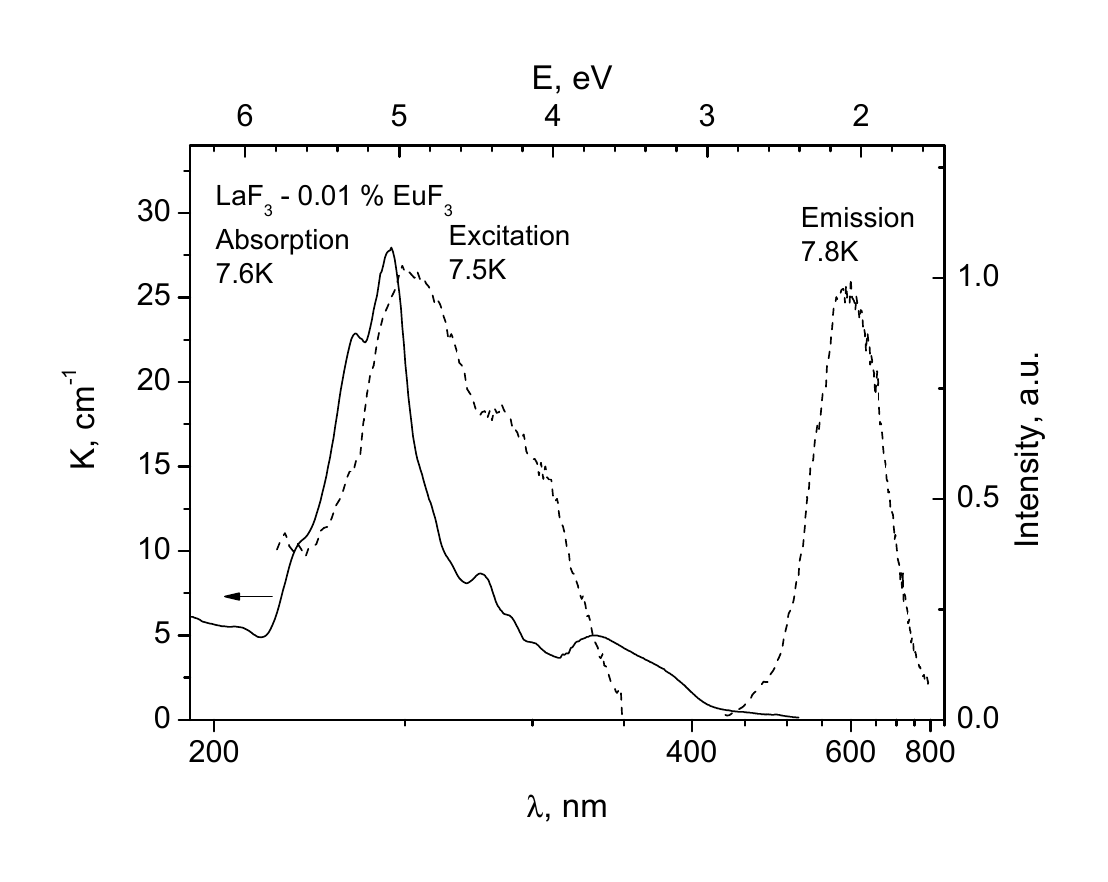}
\caption{\label{Eu_abs_ems} Absorption spectrum (full curve), excitation and emission (dashed curves) of LaF$_3$-0.01 mol.\% EuF$_3$ at shown temperatures. Excitation was measured for emission at 580 nm, the emission was measured for excitation at 270 nm.}
\end{figure}

Absorption spectrum of LaF$_3$-0.01 mol.\% EuF$_3$ contains intensive asymmetric band at 245 nm with unresolved structure and weaker long wavelength bands at 330, 380 nm (Fig.\ref{Eu_abs_ems}).  With increasing of EuF$_3$ doping the ultraviolet absorption becomes larger and at concentration near one percent of EuF$_3$ the crystal LaF$_3$ becomes yellow, due to absorption tail above 400 nm (see Fig.\ref{Eu_abs_ems}). The shape of aborption spectra does not depend on concentration of europium up to 0.3 mol.\%. At higher concentration the absorption near 245 nm becomes too large. Therefore the long-wavelength bands at 300-400 nm region belong to Eu$^{2+}$ also. Absorption band at 245 nm was ascribed to Eu$^{2+}$ ions \citep{heaps1976vacuum}. The authors have measured absorption spectrum up to 300 nm, which prevents observation of the long wavelength bands at 330, 380 nm, which also belong to europium Eu$^{2+}$ absorption. 

Red luminescence band at 600 nm was observed at low temperature. Excitation spectrum (see Fig.\ref{Eu_abs_ems}) generally correlates with 245 nm absorption bands. However red luminescence was not observed with excitation into Eu$^{2+}$ long wavelength bands (see Fig.\ref{Eu_abs_ems}). 

\begin{figure} [h]
\includegraphics[width=21pc]{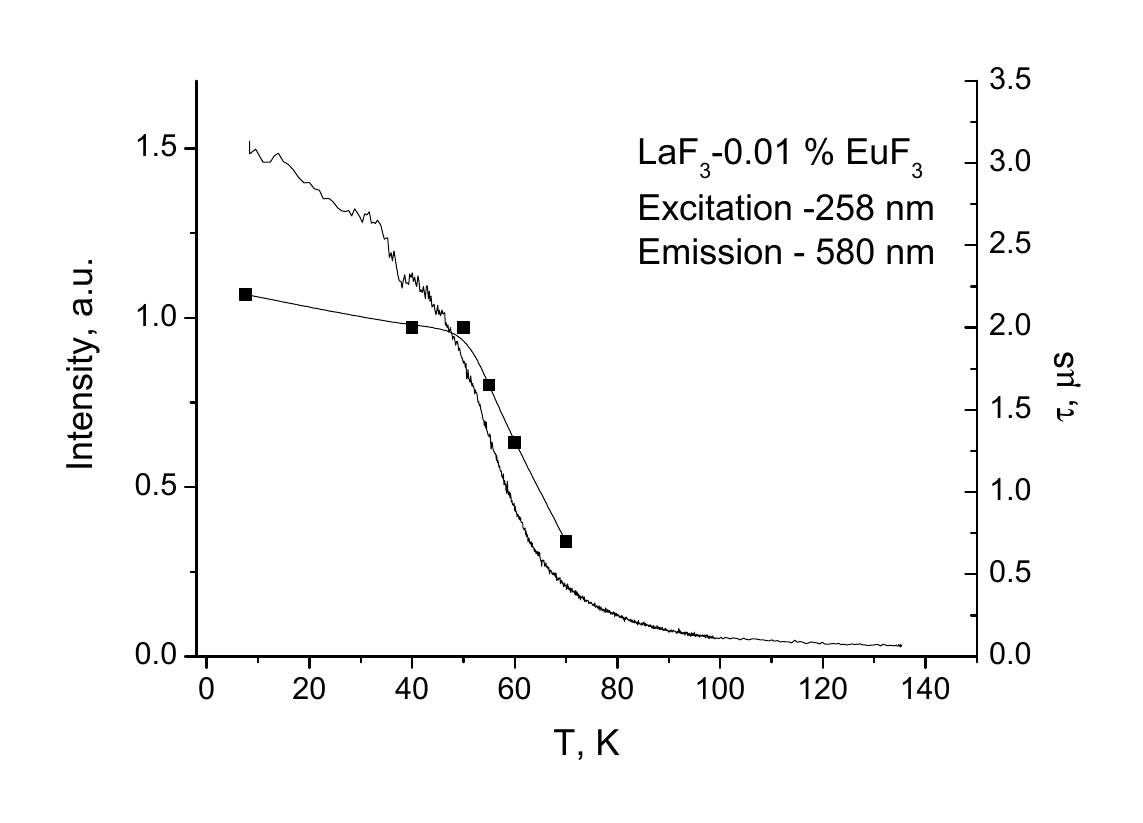}
\caption{\label{Eu-temper} Temperature dependence of intensity of Eu luminescence (full curve) and decay time (dots) of LaF$_3$-0.01 \%  EuF$_3$ crystal.}
\end{figure}

With increasing temperature the intensity of luminescence sharply decreases above 40 K (Fig.\ref{Eu-temper}). The decay time of red luminescence was 2.2 $\mu$s at 7.5 K. Above 50 K decay time sharply shortened similar to luminescence intensity (see Fig.\ref{Eu-temper}). 

\begin{figure} [h]
\includegraphics[width=21pc]{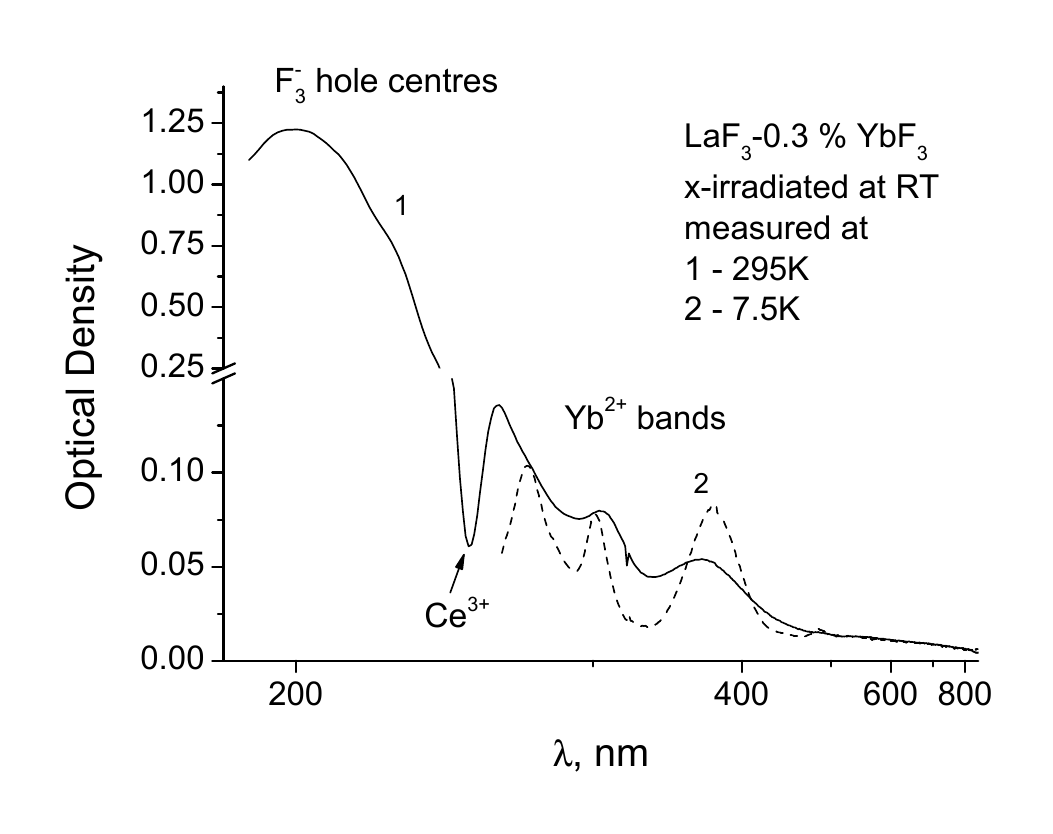}
\caption{\label{Yb-abs} Absorption spectrum of LaF$_3$-0.3 wt.\% YbF$_3$ at shown temperatures created by x-irradiation at 295K. LaF$_3$ crystal also contains unwanted Ce$^{3+}$, which absorbed at 245 nm and partially transformed under x-irradiation.}
\end{figure}

Next most probable divalent lanthanide in LaF$_3$ is ytterbium. The Yb$^{2+}$ long-wavelength bands were observed around 360, 310 nm in alkaline-earth fluoride crystals \citep{moine1989luminescence}. The Yb$^{3+}$ in LaF$_3$ shows infrared absorption near 970 nm \citep{rast1967fluorescence}. No Yb$^{2+}$ ultraviolet bands were observed in our LaF$_3$-YbF$_3$, while Yb$^{3+}$ infrared bands were grown with concentration of YbF$_3$. After x-ray irradiation of LaF$_3$-YbF$_3$ at room temperature the absorption bands at 270, 300 and 376 nm were appeared (Fig.\ref{Yb-abs}) and increased with increasing YbF$_3$ concentration. Additionally a very intensive absorption band at 200 nm, belonging to stable at room temperature  F$_3^-$ hole defects \citep{radzhabov2016}, appeared. Evidently the bands at 270, 300 and 376 nm belong to Yb$^{2+}$ in LaF$_3$.

No luminescence in the range 400-1200 nm, which can be associated with Yb$^{2+}$, was found in x-irradiated LaF$_3$-0.3 \% YbF$_3$ at temperatures down to 7.5 K. 

\subsection{Dielectric relaxation}

\begin{figure} [h]
\includegraphics[width=21pc]{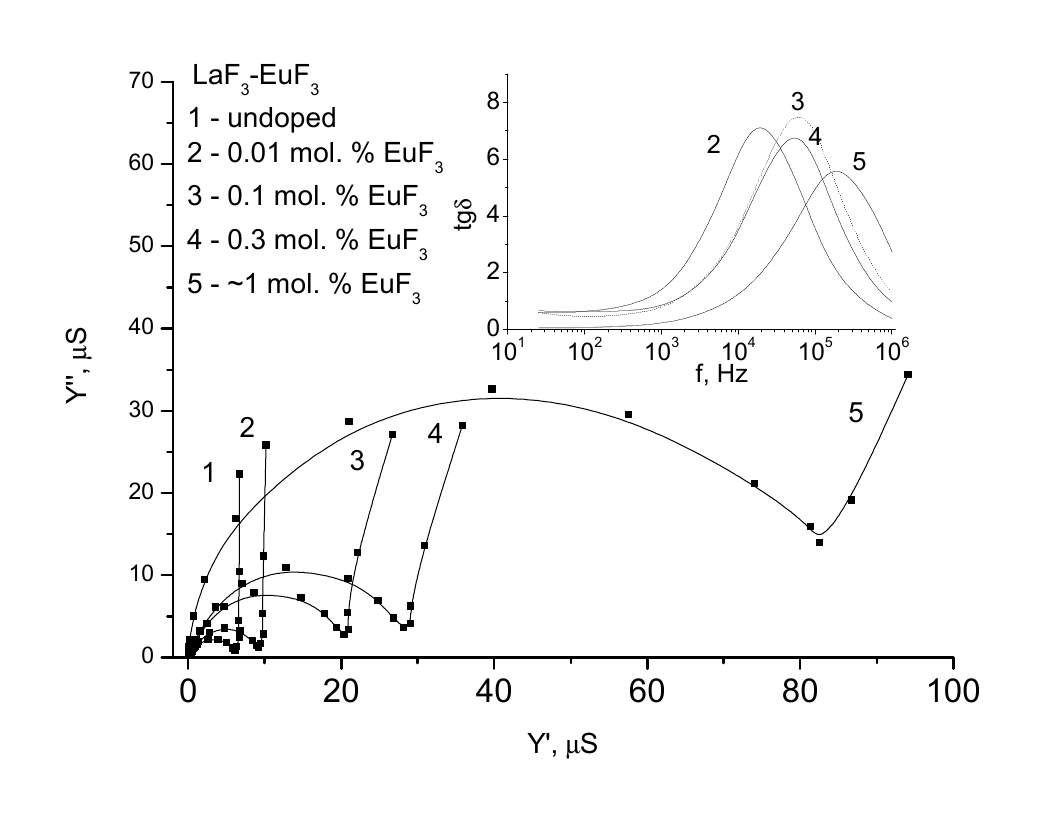}
\caption{\label{Eu-cond} Complex admittance plot (Y$^*$=G$_p$+i$\omega$C$_p$) for LaF$_3$-EuF$_3$ crystals at 295 K. Frequency range 25Hz-1 MHz. Inset shows frequency dependence of tg$\delta$ for the same samples.}
\end{figure}

Fig.\ref{Eu-cond}  presents examples of admittance plots in the complex-plane representation for cells with LaF$_3$ crystals at 295 K. The admittance plots for LaF$_3$-EuF$_3$ crystals show that the high-frequency data require an equivalent circuit composed of a frequency-independent (bulk) capacitance, Cp, in parallel with a frequency-independent (bulk) resistance, Rp. At zero point frequency is equal 25 Hz and increase till 1 MHz for last point each curve (see Fig.\ref{Eu-cond}). The high-frequency interceptions with the real axis represents the true bulk conductances.  

The increasing of europium concentration accompanied by the growth of resistive bulk conductances (see Fig.\ref{Eu-cond}) and the increasing of the Eu$^{2+}$ absorption.  

Absence of low frequency wing of tg$\delta$ (see Fig.\ref{Eu-cond}) points on absence of steady electrical conduction of LaF$_3$-Eu. It means that all of fluorine vacancies are attached to divalent europium at room temperature.  

\begin{figure} [h]
\includegraphics[width=21pc]{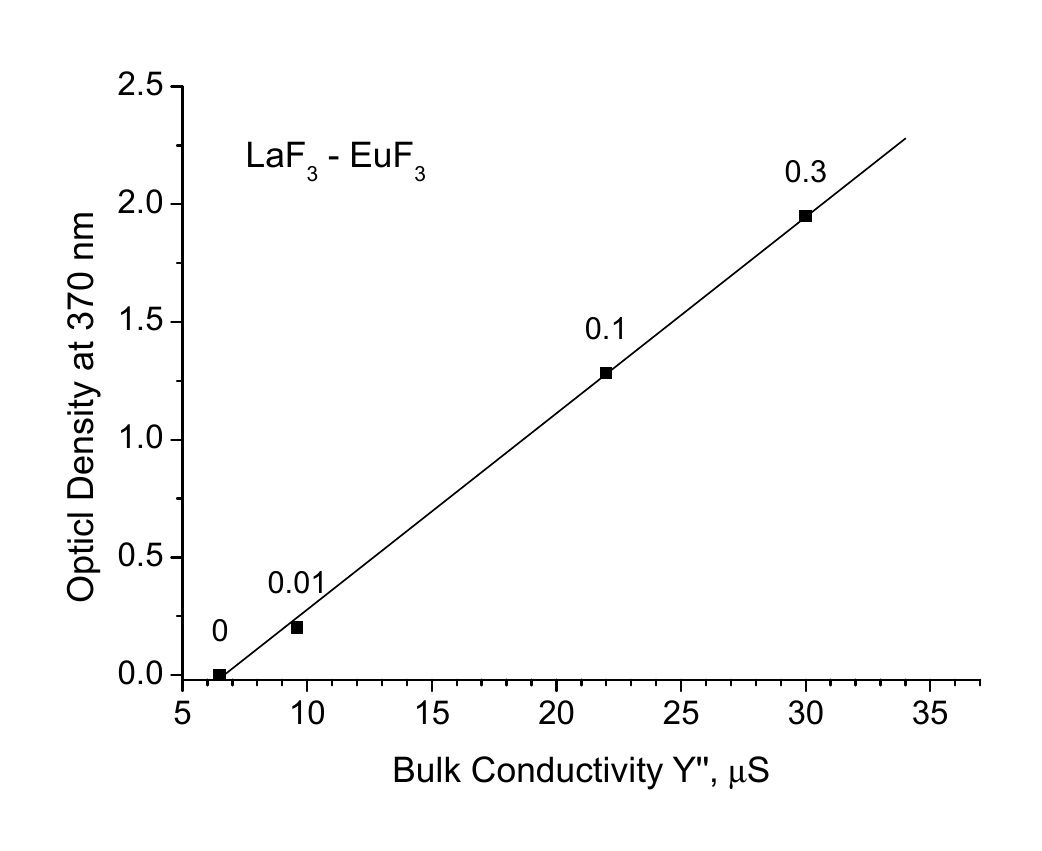}
\caption{\label{Eu-abs-cond} Bulk conductance against Eu$^{2+}$ absorption at 370 nm of LaF$_3$ and LaF$_3$-EuF$_3$ crystals.}
\end{figure}

We compare the absorption and bulk conductances of LaF$_3$ and LaF$_3$-EuF$_3$ (Fig.\ref{Eu-abs-cond}). The measurements of absorption and conductances were done on the same samples to diminish possible errors. The linear dependence was observed up to 0.3 mol. \% of EuF$_3$ dopant. Undoubtedly, the conductance of LaF$_3$-EuF$_3$ (as well as LaF$_3$-BaF$_2$ \citep{roos1985dielectric, schoonman1980solid}) are due to fluorine vacancies introduced by divalent impurity. 


\section{Discussion}

The investigations of LaF$_3$ doped with divalent alkaline-earth ions Ca$^{2+}$, Sr$^{2+}$ and Ba$^{2+}$ were proved that the charge compensators are fluorine vacancies \citep{igel1982electrical, roos1985dielectric, privalov1994nuclear}. Introduction of divalent ions into LaF$_3$ led to increasing of ionic conductivity \citep{roos1985dielectric}, appearing the peaks of thermostimulated depolarisation \citep{roos1985dielectric} and peaks of nuclear magnetic resonance of $^{19}$F \citep{privalov1994nuclear}. With increasing the Ba$^{2+}$ concentration up to 8 \% the conductivity monotonically increased \citep{roos1984ionic}. All these phenomena caused by migration of fluorine vacancies. Based on these results one could assume that charge compensator of divalent samarium is fluorine vacancy, concentration of which can be evaluated by conductivity measurements. 

Both absorption and conductivity followed the logarithmic-type growth with increasing europium doping. Finally, we obtained linear increase of LaF$_3$ conductivity with increasing Eu$^{2+}$ absorption (see Fig.\ref{Eu-abs-cond}). The linear dependence on Fig.\ref{Eu-abs-cond} are plotted using absorption at 370 nm, straight lines can be obtained for any absorptions within 200-400 nm range, also.  These results are proved that anion vacancy accompanied each divalent europium ion. Based on ionic thermodepolarisation \citep{roos1985dielectric} and dielectric relaxation investigations of Me$^{2+}$ doped LaF$_3$ \citep{roos1985dielectric, roos1984small} one could infer that anion vacancy should be in close vicinity of divalent europium.

Optical spectra of LaF$_3$-Eu$^{2+}$ has similarities with spectra of LaF$_3$-Sm$^{2+}$, investigated in our previous paper \citep{radzhabov2015spectra}. Indeed, the most intensive absorption bands in both cases belong to 4f$^n$-4f$^{n-1}$5d$^1$ transitions in divalent Sm or Eu. The weaker long-wave absorption bands, in which the emission was not excited, are present in both cases. In the case of Sm$^{2+}$ we attributed this absorption bands at 600 nm to transitions from 4f ground state of samarium to level of nearest anion vacancy. Preliminary unempirical calculations supported this conclusion \citep{radzhabov2015spectra}. Following to this the Eu$^{2+}$ bands at 330, 380 nm  can be attributed to transitions 4f$^7$ - vacancy.  Unempirical calculations, which are in progress now, will explain the detail of optical spectra. The absence of emissions after excitation into long-wave bands obviously related with large lattice relaxation around vacancy trapped electron. 

Apart from Eu no Yb luminescence was found in LaF$_3$ crystals. According to this no Yb$^{2+}$ anomalous luminescence was found in BaF$_2$, while Yb$^{2+}$ luminescence in CaF$_2$, SrF$_2$ was observed \citep{moine1989luminescence, pedrini2007photoionization}. It seems the absence of Yb$^{2+}$ luminescence related with larger lattice relaxation in excited state due to smaller ionic radius of Yb against that of Eu.

 According to Dorenbos empirical model the 4f-5d transitions of Eu$^{2+}$ in LaF$_3$ should begin at 330 nm and 5d levels fall into the conduction band \citep{dorenbos2013review}. The absence of 4f$^{n-1}$5d$^1$ - 4f$^n$ emission of divalent europium ions and absence of fine structure of 245 nm absorption 4f$^7$ - 4f$^6$5d$^1$  band correlates with the fact that 5d level fall into conduction band.

\section*{Conclusion}

Experimental results lead us to following conclusions: 

- Yb$^{2+}$ has absorption bands at 270, 300 and 376 nm, 

- each Eu$^{2+}$ is accompanied by fluorine vacancy at room temperature, this leads to appearance of long-wave absorption bands 330, 380 nm and dielectric relaxation peak,

- the Eu$^{2+}$ broadband luminescence in LaF$_3$ at 600 nm is an emission from relaxed conduction band states to ground Eu 4f level (anomalous emission).

\section*{Acknowledgments}
The authors gratefully acknowledge O. N. Solomein and O. V. Kozlovskii for preparation the crystals investigated in this work. The work was partially supported by RFBR grant 15-02-06666a.

In this work authors used the equipment of the Baikal Analytical Center for Collective Use, Siberian Branch, Russian Academy of Sciences.

\label{}





\bibliographystyle{elsarticle-harv}
\bibliography{Holmium2015}







\end{document}